\documentstyle[12pt,epsf]{article}
\begin{document}

\renewcommand{\thesection}{\Roman{section}.}
\renewcommand{\thesubsection}{\Alph{subsection}.}

\def\dxy{$d_{x^{2}-y^{2}}$ \,}
\def\dx {$d_{xy}$ \,}
\def\id{$id_{xy}$ \,}
\def\vr{$\vec{r}$ \,}
\def\vrp{$\vec{r}\prime$ \,}

%\font\mvs=mvtmsr scaled 1400 bold2
%\font\vs=mvtmsr scaled 1000 bold2

\begin{center}
{\large \bf Magnetic Field Induced $id_{xy}$ Order in a
$d_{x^{2} - y^{2}}$ Superconductor}$^\dagger$\\*[0.5in]
{\large T V Ramakrishnan}$^{*}$\\
{\bf Department of Physics\\
Indian Institute of Science\\
Bangalore 560 012, India}
\end{center}

{\large Abstract}

The interaction between planar quasiparticles in a \dxy
superconductor and quantized vortices associated with a magnetic
field perpendicular to the plane is shown to induce a pair
potential with \dx symmetry, out of phase with \dxy order. A
microscopic calculation of a process involving quasiparticle
scattering by the supercurrent around a vortex and Andreev
reflection from its core is presented. Other processes also
leading to an \id pair potential are discussed. It is
argued that such a fully gapped state may be the high field low
temperature phase observed by Krishana, Ong et al in magnetothermal
conductivity measurements of superconducting single crystal
$Bi-2212$.

\section{\bf Introduction}

Recent measurements of in-plane thermal conductivity $K_T$ in
superconducting single crystal $Bi-2212$ (1,2) for $T\ll\,T_{c}$
and $H \ll H_{c2}$ show a rather sharp change in its magnetic
field dependence at a field $H_o$ (of order a Tesla or more)
which depends on temperature $T_{o}(\sim\,10 K)$ nearly
quadratically. For $H\,>\,H_{o}$ and $T\,<\,T_{o},\,\,K_{T}$
hardly changes with the magnetic field, in sharp contrast to the
decrease exhibited below $H_o$. This suggests (1,2) a transition
from a \dxy phase to a fully gapped phase at $H_{o}(T_{o})$ with
an exponentially small density of heat current carrying
quasiparticles. I show here that in a magnetic field, a \dxy
superconductor necessarily develops a pair amplitude of \id
symmetry at low enough temperature, leading to a gapped phase
which might be the basis of the observed $K_{T}$ behaviour.

In the superconducting state, which has vortices due to the external
magnetic field, the gap quasiparticles interact with the
circulating supercurrent around each vortex, as well as with the order
parameter inhomogeneity associated with the vortex core. These
interaction terms are obtained explicitly (Section II). The
first scatters quasiparticles, eg. it changes their phase. The
second causes Andreev reflection. Due to their combined effect,
a particle interacting with a vortex comes out as a phase
shifted hole, i.e. the vortex is a source of out of phase pair
potential. This term is  calculated  in Section III, and is shown 
to have \id
symmetry. Thus, at each vortex, \id order is necessarily
induced. If the vortices are well ordered spatially, there is a
homogeneous $\Delta_{xy}$ term in the free energy, proportional
to the density of vortices  or the 
magnetic field $B$. I argue (Section IV) that this could
be the phase suggested by the observations of Krishana et al
(1,2). Related results and questions, such as the            
temperature and field scale of the transition,  effect of vortex
lattice periodicity are also discussed
(Section IV). Recently, Laughlin has proposed (3) that in a
magnetic field, the pair potential is of the form \dxy + $\alpha$ \id
where $\alpha$ is real,  and
that there is a (first order) transition to such a phase in the
$(H,T)$ plane. The proposal is based on a mapping of a \dxy +
$\alpha$ \id  Hamiltonian  to a lattice quantum Hall system, 
and on exploiting 
quantum Hall current ideas.
Here, I describe a detailed microscopic mechanism
for \id order, and
calculate its size.
           
\section{Quasiparticle Vortex Interaction}

I show here that the quasiparticle Hamiltonian in the mixed
state can be written as a sum of two classes of terms; one is
that of the uniform superconductor, and the other describes the
effect of vortices on quasiparticles. Since quasiparticles are
well defined in the superconducting state, a mean field
Hamiltonian is adequate. In the homogeneous superconductor, the
otherwise free electrons move in a real, spatially homogeneous
pair potential $\tilde{\Delta}$ with \dxy symmetry, eg.
with $\tilde{\Delta}_{k} =\Delta_{o}\,(cos\, k_{x}a - cos k_{y}a)$ for
nearest neighbour pairing. In the presence of vortices, the pair
potential $\Delta (\vec{r},\vec{r\prime})$ is inhomogeneous and
complex. Its phase changes by $2\pi$ on going round a vortex,
and the magnitude must vanish at each vortex core
$\vec{R}_{\ell}$. Separating out the phase part, we can write
$\Delta (\vec{r},\vec{r\prime})$ as

$$\Delta\,(\vec{r}\vec{r}\prime) =
\Delta_{m}(\vec{r}-\vec{r}\prime, {\vec{R}-\vec{R}_{\ell}}\,)\,
exp\,[\,\frac{i}{2}\{\,\sum_{\ell}\,\theta\,(\vec{r}-\vec{R}_{\ell})
+\theta\,(\vec{r}\prime-\vec{R}_{\ell})\}\,].$$ 

Here $\vec{R}$ is 
the centre of mass coordinate ${\,(\vec{r}+\vec{r\prime})/2}$.
Making a
gauge transformation $\tilde{\psi}^{+}_{\uparrow}(\vec{r})\,
exp \{\,\frac{i}{2}\sum_{\ell}\theta(\vec{r}-\vec{R}_{\ell)}\}\,
\Rightarrow\psi^{+}_{\uparrow}(\vec{r})$, 
the quasiparticle Hamiltonian becomes

\begin{eqnarray}
H &=&\frac{1}{2m}
\sum_{\sigma}\int\,d\vec{r}\,\psi^{+}_{\sigma}(\vec{r})\,(\vec{p}+\,
m\vec{v}_{s}(\vec{r})\,)^{2}\,\psi_{\sigma}(\vec{r})\\
&&+\int d\vec{r}\,d\vec{r\prime} [\,\Delta_{m}
(\vec{r}-\vec{r\prime},\{\,{\vec{R}-\vec{R}_{\ell}\}\,)\psi^{+}_{\uparrow}
(\vec{r})\psi^{+}_{\downarrow}(\vec{r\prime})} + h.c.]\,\,\,(1)
\end{eqnarray}

where 
$$\vec{v}_{s}(\vec{r})
=(2m)^{-1}\sum_{\ell}\,\{\hbar\vec{\bigtriangledown}\theta\,(\vec{r}-
\vec{R}_{\ell})\,-2e\vec{A}(\vec{r}-\vec{R}_{\ell})/c \}$$

is the in-plane gauge invariant superfluid velocity. In the
London limit, $\Delta_{m}$ can be considered spatially uniform
($\vec{R}$ independent) with \dxy symmetry, except
for order parameter deficits $\delta\Delta_{m}$ around each
vortex. (These have to be determined self consistently)

The quasiparticle Hamiltonian Eq.(1) can be written as the sum
of an unperturbed term $H_{o}$, and the remainder describing
quasiparticle vortex interaction. In the momentum
representation, and the Nambu formalism, one has 
$$H = H_{o} \,
+\,(H_{\theta} + H_{m} + H_{KE}\,)\eqno(4a)$$ 
where 
$$ H_{o} =\sum_{k}a^{\dagger}_{k} \,(\tilde{\epsilon}_{k}\,\tau_{3} +\,
\Delta_{k}\tau_{1}\,)a_{k}\eqno(4b)$$ 
$$H_{\theta}\,=\,
\sum_{k,q}a^{\dagger}_{k}\,{\hbar\vec{k}.\vec{v}_{s}(\vec{q})\,}a_{k-q}\eqno(4c)$$
$$H_{m} = -\Delta_{o}\sum_{k,q}f_{kq}(a^{\dagger}_{k}\tau_{1}\,
a_{k-q}\,)\eqno(4d)$$ 
and $H_{KE}$ is the superfluid kinetic
energy. Eq.(4b) describes a $d_{x^{2}-y^{2}}$ superconductor.
$H_{\theta}$ is the quasiparticle superfluid velocity
interaction and $H_m$ is the inhomogeneous pair potential due to
the vortex core, with $-\Delta_{o} f_{\vec{k},\vec{q}}$ being the
Fourier transform of the deficit
$\delta\Delta_{m}(\vec{r}-\vec{r}\prime,\{\vec{R}-\vec{R}_{\ell}\}\,)$.
The superfluid velocity $\vec{v}_{s}(\vec{q})$ is assumed to
have the standard Ginzburg Landau form 
$$\vec{v}_{s}(\vec{q}) =
\frac{\lambda^{2}}{2m}\,\frac{1}{2A}\,\sum_{\vec{q}\ell} 
\left(\frac{i\vec{q}\times\,\hat{e}_{z}}{1+\lambda^{2}q^{2}}\right) 
e^{i\vec{q}.\vec{R}_{\ell}}\,\eqno(5)$$ 
away from the vortex cores. For
wavevector transfers $q\ll\lambda^{-1}$, this has the unscreened
form independent of $\lambda^{2}$. We now use the mean field
Hamiltonian Eq.(4) to show that $id_{xy}$ order is induced in a
magnetic field.

\section{$id_{xy}$ Order}

The form Eq.(4) is natural for looking into the question of the
pair potential in the 
presence of a magnetic field. In its absence, $H_{\theta},
H_{m}$ and $H_{KE}$ vanish, and $\Delta_{k}$ in Eq.(4b) is just
the $d_{x^{2}-y^{2}}$ uniform value. We imagine $H_{o}$ and
$H_{m}$ being turned on, and ask if any out of phase order is
induced, i.e., in the notation above, whether
$\lambda_{k}=\langle a^{\dagger}_{k}\,\tau_{2}\,a_{k}\rangle \neq 0$.
We are also interested in the $k$ dependence of $\lambda_{k}$.
We shall calculate $\lambda_{k}$ with $H_{\theta}$ and $H_{m}$
as perturbations. This enables us to focus directly on processes
leading to a particular order parameter symmetry.

The first nonvanishing contribution to $\lambda_{k}$ comes from
a term linear in $H_{\theta}$ as well as in $H_{m}$; there are
no contributions to second order, either in $H_{\theta}$ or
$H_m$. The process is diagrammatically represented in Fig.1a. It
describes the combined effect of quasiparticle scattering from
state $\vec{k}$ to $(\vec{k}-\vec{q})$ by the supercurrent due to the
vortex, and
Andreev reflection from ($\vec{k}-\vec{q}$) to a hole of
momentum-$\vec{k}$ caused
by the inhomogeneous order parameter magnitude associated with
the vortex core.

The pair amplitude $\lambda_{k}$ can be calculated from
Fig.(1a), and Eq.(4). At $T=0$, it is
$$\langle a^{\dagger}_{k}\tau_{2}a_{k}\rangle = \sum_{q}(\Delta_{o}
f_{kq}) \{\,(\hbar^{2}/2mA)\, (\vec{k}\times\,i\vec{q}.\hat{e}_{z}\,)q^{-2}\}\,
[(\epsilon_{k-q}/E_{k}E_{k-q}(E_{k} + E_{k-q})].\eqno(6)$$

In Eq.(6), the first bracket is the pair potential due to
the vortex core; approximately,
$f_{kq}= f_{q}\{cos(k_{x}+q_{x})a-cos(k_{y}+q_{y})a\}$
where $\sum_{q}\,f_{q} =1$.
The
second factor, in curly brackets, is the quasiparticle
supercurrent coupling $\vec{k}.\vec{v}_{s}(\vec{q})$ for $q\gg
\lambda^{-1}$ (the regime of interest). The last term arises 
from the intermediate state sum, with 
$E_{k} =|(\epsilon_{k}^{2} + \Delta^{2}_{k})^{1/2}|$ . On
expanding the integrand as a power series in $q$, the leading
term is 
$$\lambda_{k} =
\alpha\,(n_{v}/n)\,(\Delta_{o}/E_{k})\,
(\hat{k}_{x}\hat{k}_{y})\eqno(7)$$

Here, $\alpha$ is a constant of order unity, and $n_{v}$ is the
vortex density.

We notice that $\lambda_{k}$ has $d_{xy}$ symmetry. It is a result of the
coupling between the internal and centre of mass states of the pair
near a vortex. In the Fourier representation used, the former is
described by the momentum $\vec{k}$, and the latter by $\vec{q}$.
The quasiparticle supercurrent coupling has a structure
$(\vec{k}\times\,\vec{q}).\hat{e}_{z}$ that affects the angular state
of a pair, through the constituent single  particle dispersion.
Algebraically, $(\vec{k}\times\vec{q}).\hat{e}_{z}\,(\epsilon_{\vec{k}-\vec{q}})
\simeq\,k_{x}k_{y}\,(q^{2}_{x}-q^{2}_{y})$. This term is clearly  nonlocal, ie. it
arises from higher powers of $q$ or of gradient/derivative terms.
Thus, a fully local, semiclassical theory, working in terms of a spatially slowly
varying superfluid velocity $\vec{v}_{s}(\vec{r})$ and a local (diagonal)
quasiparticle momentum or Fermi  surface (4), will miss this effect, while it
may be appropriate for quasiparticle density of states etc.. The $id_{xy}$ order can
also be induced by higher order terms involving only $H_{\theta}$ or the
quasiparticle-supercurrent interaction. An example is shown in Fig.(1b). Other
contributions, involving induced $s$ wave order, are also possible.

We have focussed so far on spatially uniform single site terms. The $id_{xy}$
order induced around each vortex is spatially nonuniform. The leading nonvanishing
contribution is of ${\it first\, order}$ in $H_{\theta}$, and is shown in Fig.1c.
It describes a $d_{x^{2}-y^{2}}$ pair becoming a $d_{xy}$ pair {\it and} simultaneously
 acquiring a momentum $q$,  near a vortex. For $q\gg\lambda^{-1}$, the contribution from this
 diagram is nonvanishing at small $q$. However, because  of London screening, the
 $q\ll\lambda^{-1}$ limit vanishes. A related process involving $d_{xy}$ order near a
 spin orbit impurity potential has been recently discussed by Balatsky (5).

 The perturbative approach described above raises the obvious question of the 
 expansion parameter.
 On general coupling constant and phase space grounds this can be argued to be
 $(1/k_{F}\xi)$ for both $H_{\theta}$ and $H_{m}$, where $\xi$ is the pair coherence length. 
This is about (1/5) or (1/6). Thus, perturbation theory is expected to converge.

 The $id_{xy}$ order $\lambda_{k}$ leads to a gap $i\Delta_{xy}$ with $xy$ symmetry,
 in two ways. A process for $\lambda_{k}$ corresponding to Fig.1a has, associated
 with it, an anomalous self energy Fig.1d, which leads to an $xy$ symmetry gap
 function. Thus even if there is no interaction in the $xy$ particle particle channel,
 a $\Delta_{xy}$ is induced. Secondly, there may actually be such a potential (attractive
 or repulsive), say $V_{xy}$. Then in the mean field approximation, a gap
 $\Delta_{xy}(k)\,\simeq\,\sum_{k\prime}V_{xy}\,(kk^{\prime})\,\lambda_{k^\prime}$
 is induced. The question of self consistency can be fully addressed only if the microscopic
 mechanism of pairing (say in the $d_{x^{2}-y^{2}}$
 channel) is known. In a BCS theory, if appropriate pseudopotentials $V_{x^{2}-y^{2}}$
 and $V_{xy}$ are assumed, the relevant
 Gor'kov equations (in the presence of vortices) or the quasiparticle problem
 (described by Bogoliuov-de Gennes equations) need to be solved self consistently.
 We have not done this; the lack of self  consistency does not affect either the existence
 or the approximate size  of the $T=0\, \,id_{xy}$ pair amplitude.

 \section{The Gapped Phase}

 We have shown above that an $id_{xy}$ pair amplitude is inevitable for a
\dxy superconductor in a magnetic field, and that there is a consequent
 $id_{xy}$ gap. Thus, the ground state of the system in a magnetic field is necessarily
 a fully gapped superconductor in which the gap parameter
 $\Delta_{k}=\sqrt{\Delta^{2}_{x^{2}-y^{2}}(k) +\Delta^{2}_{xy}(k)}$ is
 nonzero at all points $k$ on the Fermi surface. This has obvious thermodynamic and
 transport consequences. In particular, since the number of quasiparticles
 is exponentially small at very low temperatures, their contribution
  to thermal conductivity is negligible. We thus expect, at low enough temperature
  and sufficient vortex  density, that there will be a transition to a gapped phase. We
  do not have a complete theory of this transition, but discuss possibilities
  below, following a brief analysis of experimental results (1,2).

  As mentioned earlier, Krishana et al (1,2) find a transition at $T_o$ from a $K_{T}$ which 
  decreases with increasing $H$ to a field independent value at $H_o$; approximately,
  $T_{o}\propto\sqrt{H_{o}}$.
  It is argued
  that this implies a thermodynamic transition to a gapped or coherent phase
  from an ungapped, incoherent phase. The  transition is most likely continuous,
  since for a discontinuous change one expects a jump in $K_{T}$ at the transition
  point.

  The physical picture of the transition is that around each vortex an inhomogeneous
  $d_{xy}$ order develops. If the vortices are not regularly arranged, and if the
  $d_{xy}$ amplitudes are small as well as patchy or disconnected, quasiparticles
  see an inhomogeneous medium which scatters them, and a pseudogap develops near the nodes;
  this pseudogap deepens as the vortex density increases so that the electronic
  thermal conductivity decreases with increasing vortex density. At some critical
  vortex  density dependent on temperature, the $id_{xy}$ order parameters
  overlap, or the vortices order spatially, and a nonzero gap develops for the lowest
  lying quasiparticle excitations. This is the new phase.

If the $d_{xy}$ pair amplitude, (labelled $m^{k}_{xy}$) is the
expected order parameter, an order parameter functional can be
obtained using the standard auxiliary field method. We have
calculated low order terms at $T=0$ in such an expansion. There is
a linear term, due to the fact that averages $\langle
a^{\dagger}_{k}\tau_{2}a_{k}\rangle$ with $k_{x}k_{y}$ symmetry,
discussed in Section III, are nonzero. There are quadratic terms
arising from both $V_{xy}$ symmetry interactions between
electrons (pairs) as well as a nonzero $xy$ pair susceptibility.
There is a cubic kinetic energy term (3). Because of the linear
term, $m^{k}_{xy}$ is nonzero;
unsurprisingly, it has the value $\langle
a^{\dagger}_{k}\tau_{2}a_{k}\rangle_{xy}$ for which some terms have
been indicated in Section III. We have not carried through the
calculation for $T\neq 0$, and so are unable to find the
temperature below which $m^{k}_{xy}$ becomes nonzero. The linear
in $m$ term persists however at $T\neq 0$.

The gap in the excitation spectrum at $T=0$ is due to both 
anomalous self energy (Fig. 1d) and the
mean $xy$ pair potential if there is a $V_{xy}$ interaction. The
gap due to the former is approximately $\Delta_{o}(n_{v}/n)$ and
due to the latter,
$4\,(|V_{xy}|/\epsilon_{F})\,\Delta_{o}(n_{v}/n)$. Here
$\Delta_{o}$ is the $d_{xy}$ gap magnitude and $n$ is the
carrier density. The actual value thus depends on $V_{xy}$ which
is not known. Assuming $V_{xy}/\epsilon_{F}\,\simeq\,1$ (a
large value) and $\Delta_{o}\simeq\,300K$, for a field of 5T, 
the minimum gap $\Delta_{xy}$
is about 10K. This comparable to the
temperature 20K at which the gapped phase transition occurs
for 5T, though smaller.
Also, equating $T_o$ to $\Delta_{xy}$, we find that $T_{o} \propto
H$; the experimental points are closer to $\sqrt{H}$.

This suggests that the identification of the observed transition
with the development of a uniform part of $d_{xy}$ pair
amplitude may not be correct. One curious feature of this
identification is that even when the vortices are distributed
randomly the pair amplitude is uniform (there is a
diagonal in $k$ term)
It is the sum of $N_{v}$ independent identical single
vortex terms irrespective of their arrangement. 
However, the $d_{xy}$ order developed due to
quasiparticle vortex interaction is spatially inhomogeneous, on
the scale of the screening length $\lambda$, as discussed in
Section III. If the vortices order spatially,
there are coherent
terms linear in $H_{\theta}$, with wavevectors $\vec{q}$ equal to
Bragg vectors
of the reciprocal vortex lattice. (These are $\vec{q}
=\vec{G}$ components of the term shown in Fig.1c\/). They have
the right energy scale.
The relation between the magnetic field and the related energy
scale is $T_{c}\propto\,G$ or $T_{c}\propto \sqrt{H}$, close to
that observed. The gapped nature of this vortex lattice phase,
the preferred lattice structure, and the nature of the
transition to this lattice on cooling, are being investigated
(6).

{\bf Acknowledgements:}

I would like to thank K Krishana and N P Ong for sharing their
experimental results, and the Department of Science and
Technology, New Delhi, for partial travel support.

{\bf References}

\begin{itemize}
\item[$\dagger$] Invited talk at the conference {\em Spectroscopies of Novel
Superconductors}, SNS '97 at Cape Cod, Sept. 1997
\item[*] Also at Jawaharlal Nehru Center for Advanced Scientific Research,
Bangalore-64
\item[1.] K Krishana et al, Science, {\bf 277}, 83 (1997).
\item[2.] N P Ong, this conference.
\item[3.] R B Laughlin, preprint, cond.-mat 9709004.
\item[4.] C K\"{u}bert and P Hirschfeld, preprint; cond.-mat 9708221.
\item[5.] A V Balatsky, this conference, and preprint cond.-mat 9709287.
\item[6.] T V Ramakrishnan, to be published.
\end{itemize}

\newpage
{\bf Figure Captions}

Diagrams describing processes leading to $id_{xy}$ pair
amplitude and related anomalous self energy.

{\bf (a)}\hskip5mm\parbox[t]{4.5in}{An interference term for
$id_{xy}$ order involving a quasiparticle (straight line) being
scattered by supercurrent (wavy line) around a vortex (marked by
a cross), and by order parameter inhomogeneity around the vortex
core (dotted line).} \\*[5mm]
{\bf (b)}\hskip5mm\parbox[t]{4.5in}{A
higher order term for $id_{xy}$ order, solely from quasiparticle
supercurrent coupling $H_{o}$.} \\*[5mm]
{\bf (c)}\hskip5mm\parbox[t]{4.5in}{A
spatially inhomogeneous $id_{xy}$ pair amplitude term, first order in 
$H_{\theta}$.} \\*[5mm]
{\bf (d)}\hskip5mm\parbox[t]{4.5in}{Anomalous self energy connected
with the process of Fig. (1a).}
\vskip 1cm
\centerline{\epsffile{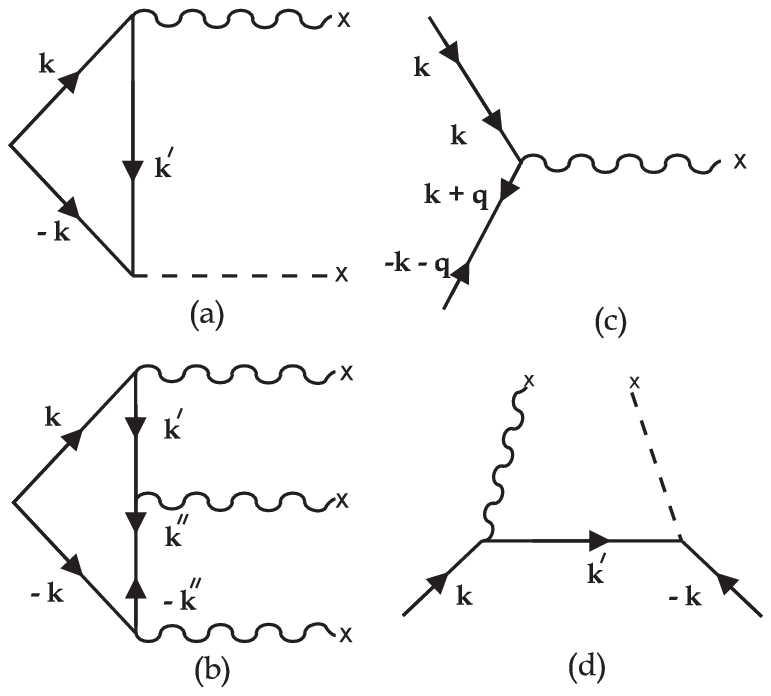}}

\end{document}